# Temperature dependence of mycosubtilin homologues production in *Bacillus subtilis* ATCC6633

Patrick Fickers [a,b], Valérie Leclère [a], Jean-Sébastien Guez [a], Max Béchet [a], Françoise Coucheney [a], Bernard Joris [b], Philippe Jacques [a*]

[a] Laboratoire de Procédés Biologiques, Génie Enzymatique et Microbien (ProBioGEM, UPRES EA 1026), Polytech'Lille, Université des Sciences et Technologies de Lille, F-59655 Villeneuve d'Ascq Cedex, France

[b] Centre d'Ingénierie des Protéines, Laboratoire de Physiologie et Génétique Bactrienne, Université de Liège, Institut de Chimie, Bat B6, B-4000 Liege, Belgique.

Patrick Fickers : pfickers@ulg.ac.be

Valérie Leclère : valerie.leclere@univ-lille1.fr

Jean-Sébastien Guez : jean-sebastien.guez@polytech-lille.fr

Max Béchet : max.bechet@univ-lille1.fr

Françoise Coucheney : Francoise.Coucheney@univ-lille1.fr

Bernard Joris : bjoris@ulg.ac.be

Philippe Jacques : philippe.jacques@polytech-lille.fr *Correspondence and reprints

**Abstract**

*Bacillus subtilis* ATCC6633 produces mycosubtilin, a non-ribosomally synthesized lipopeptide of the iturin family which presents antagonistic activities against various phytopathogens. Different homologues with fatty acid moiety varying from $C_{15}$ to $C_{17}$ are usually co-produced with their biological activities increasing with the number of carbon in the fatty acid chain. In the present report, we highlight that growth temperature modulates either the level of the mycosubtilin production and the relative abundance of the different homologues. A 30-fold increase in mycosubtilin production was observed when the temperature was decreased from 37 °C to 25 °C for both strain ATCC6633 and its derivative BBG100, a constitutive mycosubtilin overproducer. However, no significant difference in both the expression of the mycosubtilin synthetase encoding genes and in the intracellular synthetase concentration could be found, suggesting that the observed phenotype originated from a higher mycosubtilin synthetase turnover at lower temperature. We also point out that a lower growth temperature leads to an increased proportion of odd-numbered fatty acid homologues as a consequence of the *de novo* synthesis of C17 anteiso fatty acid following the cell adaptation to low temperature.

# 1. Introduction

Members of the *Bacillus subtilis* family produce a wide variety of antibacterial and antifungal antibiotics (for review see [32]). Some of them, such as bacilysin, chlorotetain, mycobacillin, difficidin and lipopeptides are formed by nonribosomal peptide synthetases and/or polyketide synthetases. The lipopeptides belonging to the surfactin, iturin and fengycin families [37] are amphiphilic cyclic peptides composed of seven α-amino acids (surfactins and iturins) or ten α-amino acids (fengycins) linked to a single β-amino fatty acid (iturins) or β-hydroxy fatty acids (surfactins and fengycins). The length of the fatty acid moiety may vary from $C_{13}$ to $C_{16}$ for surfactins, from $C_{14}$ to $C_{17}$ for iturins and from $C_{14}$ to $C_{18}$ in the case of fengycins. Different homologous compounds with a linear or branched fatty acid moiety are usually co-produced for each lipopeptide family [28].

*B. subtilis* ATCC6633 produces subtilin [31], subtilosin [30], rhizocticin [18], and two lipopeptides, surfactin and mycosubtilin, the latter being a member of the iturins family [4,5]. Mass spectrometry analyses of *B. subtilis* ATCC6633 supernatant cultured at 30 °C revealed that the two main mycosubtilins produced belong to $C_{16}$ and $C_{17}$ homologues [21]. The mycosubtilin gene cluster spans about 38 kb and consists of four ORFs designated *fenF* and *mycA, mycB* and *mycC*, all of them being under control of the *myc* promoter [4,5]. The subunits encoded by the three *myc* genes contain the seven modules necessary to synthesize the peptide moiety of mycosubtilin. They show strong similarity with members of the peptide synthetase family and display the ordered assembly of conserved condensation, adenylation, and thiolation domains. Iturins present a strong fungitoxic activity against different phytopathogens such as *Botrytis cinerea*, *Fusarium oxysporum* and *Pythium aphanidernatum*. Biological activities of iturins have been shown to increase with the number of carbon atoms in the fatty acid chain. Indeed, $C_{17}$ homologues are 20-fold more active against pathogens than the $C_{14}$ isoform [14].

Of the biological control alternatives to chemical pesticides used for reducing plant diseases, the application of non-pathogenic soil bacteria living in association with plant roots is promising. These bacteria can antagonize fungal pathogens by producing low-molecular-weight fungitoxic compounds, such as the above-mentioned lipopeptides [28]. We recently highlighted in biocontrol assays, conducted with tomato/*Pythium* pathosystem, that pre-treatment of tomato seeds with the mycosubtilin-overproducing derivative BBG100 prior to planting led to enhanced seedling emergence. This demonstrated that overproduction of mycosubtilin gained protection against *Pythium* damping-off of tomato seedlings [21].

In the present study, we demonstrate that growth temperature modulates both the level of mycosubtilin production and the relative abundance of the different homologues produced without any influence on *myc* expression and synthetase intracellular concentration.

## 2. Material and Methods

*2.1. Plasmids, strains and media*

Plasmids, primers and *B. subtilis* strains used in this study are listed in Table 1. *E. coli* strain DH5α (Promega, Madison, WI, USA), used for transformation and amplification of recombinant plasmid DNA, was grown at 37 °C in Luria-Bertani medium supplemented with ampicillin (Sigma, St Louis, MO, USA, 100 µg/ml) or chloramphenicol (Sigma, 5 µg/ml) when required. Conventional calcium-shock procedure was used for *E. coli* transformation [29] whereas *B. subtilis* was transformed as described elsewhere [15]. All cultures were performed in Landy medium [20]. α-amylase activity was detected by growing *Bacillus* colonies overnight on LB plates containing 1% soluble starch and staining plates with iodine as described elsewhere [2]. Microbial growth was monitored by dry weight (D.W.) determination.

*2.2. DNA manipulation*

Standard molecular genetic techniques were used [29]. Restriction enzymes were obtained from Fermentas GMBH (St. Leon-Rot, Germany) or New England Biolabs (Beverly, MS, USA). All primers were from Eurogentec (Seraing, Belgium). PCR products were purified with the QIAquick PCR purification kit (Qiagen, Hilden, Germany). Plasmid DNA was extracted and purified from *E. coli* with the QIAprep Spin Miniprep kit (Qiagen). DNA fragments were purified from agarose gels using the QIAquick Gel Extraction Kit (Qiagen). Arrow Taq polymerase (Qbiogene, Montreal, Canada) was used for polymerase chain reaction as recommended by the manufacturer.

*2.3. Construction of strain BBG117*

To measure the expression of the *myc* operon, a *lacZ*-reporter gene was obtained using vector pDG1661 which contains the ribosomal binding site of the *B. subtilis spoVG* gene fused to a promoter-less *lacZ* cassette [11]. The *myc* promoter fragment to be tested was obtained by PCR using the primers pMYCfo and pMYCrev (Table 1), and then subcloned into pGEM-T Easy vector to generate pBG107. This latter was then sequenced using the universal T7 and SP6 primers to verify the absence of PCR mistake. pBG107 was *Eco*RI-*Bam*HI digested and the resulting 942-bp fragment was subcloned into pDG1661 at the corresponding sites to yield pBG111. This reporter construct was then integrated into the *B. subtilis* ATCC6633 *amyE* locus, giving rise to strain BBG117. Correct integration was verified by both the loss of amylase activity and by analytical PCR using primers AmyEfo and AmyERev (Table 1).

*2.4 Construction of strain RFB136*

The *ΔmycA* cassette used to disrupt the *myc* operon was constructed as previously described [6]. First, the ~1,4 kb P and T fragment consisting to part of 3' and 5' *mycA* ORF were PCR amplified using primer pair RFO120/RFO121 and RFO122/RFO123 respectively, and *B. subtilis* ATCC 6633 genomic DNA as a template. Primers RFO121 and RFO122 contain the rare meganuclease I-*Sce*I recognition sequence. P-I*Sce*I and I*Sce*I-T fragments were then pooled and used as a template for amplification of the P-I*Sce*I-T cassette with primers RFO120 and RFO123. The resulting fragment was then cloned into pGEM-T Easy vector to generate RFP119. The 1,6 kb fragment encoding a kanamycin resistance gene was rescued from RFP104 by I*Sce*I digestion and subcloned into RFP119 at the corresponding restriction site to yield RFP120. The *ΔmycA* cassette was used to transform *B. subtilis* ATCC6633 and transformants were selected on LB kanamycin plates. Integration by double crossing event was verified by analytical PCR using the primer pair RFO120/RFO123. The absence of mycosubtilin production in culture supernatant was verified by matrix-assisted laser desorption ionization-time of flight mass spectrometry (MALDI-TOF) on a Bruker Ultreflex tof (Bruker Daltonics, Bremen, Gremany) as previously described [21]. The mycosubtilin non-producer derivative was named RFB136.

*2.4 β-galactosidase assays*

β-galactosidase activities were measured according to Fickers et al. [7] on cell extracts prepared by chloroform treatment and centrifugation. One unit of β-galactosidase activity is defined as the amount of enzyme that produces 1 nmole of o-nitrophenol min$^{-1}$ at 37°C.

*2.5. Reverse transcription-polymerase chain reaction (RT-PCR) and comparative PCR*

Total RNAs were isolated from culture cells with RNeasy Protect Bacteria Mini Kit (Quiagen), with some modifications [12]. Total RNAs were quantified on a Nanodrop ND-100 spectrophotometer whereas their integrity was estimated by calculating the RNA 23S/16S ratio on an Agilent 2100 Bioanalyser equipped with a RNA 6000 nanoLabChip. Extracted RNAs were then subjected to reverse transcription with Superscript II reverse transcriptase as recommended by the manufacturer (Invitrogen, Carlsbad, CA, USA). *fenF, rplL* and *cspB* cDNAs were PCR-amplified using the primer pairs fenFfo/fenFrev, rplLfo/rplLrev and cspBfo/cspBrev, respectively. The PCR products were then analysed by gel electrophoresis on a 1.5% agarose gels and the median-base trimmed mean density (MTM) were determined by scanning densitometry using the Arrayvision software (GE Healthcare, Upsalla, Sweden). MTM values of each DNA fragment corresponding to *fenF* and *cspB* were normalized for the level of the *B. subtilis* housekeeping gene *rplL* in the same sample.

*2.6. Lipopeptide purification and identification*

One ml culture samples were centrifuged at 13 000 x *g* before being loaded onto $C_{18}$ Maxi-Clean cartridges (1 ml bead volume, Alltech, Deerfield, IL, USA), washed successively with 8 ml of water and 8 ml of $H_2O$/methanol mixture (1:1, v/v). Lipopeptides were then eluted with 5 ml pure methanol, dried under vacuum and resuspended in 100 µl of methanol. Mycosubtilin concentration was determined by RP-HPLC using a Vydac 218 TP $C_{18}$ column (250 x 4.6 mm, 5 µM packing, Vydac, Hesperia, CA, USA). The mobile phase was an acetonitrile/$H_2O$/trifluoroacetic acid mixture (40:60:0,5, v/v/v). Samples (20 µl) were eluted at a flow rate of 1 ml $min^{-1}$. Purified iturins standard were purchased from Sigma. Mycosubtilins were identified based on the second derivatives of their UV-visible spectra (Waters PDA 996 diode array; Millenium software, Milford, MA, USA). The different homologues and the fatty acid isomers (branched or linear) were determined based on their retention time as described elsewhere [11]. For normalization, mycosubtilin productions were expressed as a percentage of the highest value obtained.

*2.7. Preparation of the cell free extract and enzyme purification*

Preparation of cell free extracts was performed as described elsewhere with some modifications [36]. Briefly, cells in late exponential growth phase, cultured in 500 ml Landy medium, were resuspended in 10 ml Tris-HCl buffer 50 mM pH 7 containing 3 mM DTT, 3 mM EDTA, 2 mM benzamidine and 20 % sucrose. Protoplasts were generated by lysosyme treatment (5 mg ml$^{-1}$ cell suspension) at 30 °C for 35 min. They were then freezed at –80 °C, thawed rapidly in a 30°C water bath and finaly sonicated 3 times for 30 sec on ice. Cell debris were removed by centrifugation at 25 000 x g for 30 min. Nucleic acids were precipitated by addition of streptomycin sulfate at a final concentration of 1 % (w/v) and stirring for 10 min. They were then peletted by centrifugation at 25 000 x g for 45 min. Proteins in the supernatant were salted out with ammonium sulfate at 70 % saturation and dissolved in a minimum volume of 50 mM Tris-HCl buffer pH 7 containing 3 mM DTT, 3 mM EDTA and 10 % sucrose. Five hundred µl of the crude enzyme extract were loaded on an ultrogel AcA 34 column (Sigma, 15 x 1,5 cm) and eluted with the same buffer at a flow-rate of 0,3 ml min$^{-1}$. Fractions of 1 ml were collected.

*2.8. Methods for protein analysis*

Protein concentration was determined by the method of Bradford [1] using the Bio-Rad protein assay solution (Bio-Rad Laboratories, Hercules, CA, USA). SDS-PAGE was performed according to Laemmli [19] in a 5% (w/v) acrylamide gel at a constant current of 30 mA. The following marker proteins were used : β-galactosidase (116 kDa), apoferritin (443 kDa) and thyroglobulin (669 kDa). After electrophoresis, gels were colored using coomassie brilliant blue G-250 using standard procedures. For N-terminal sequence determination, proteins were electrobloted onto an Hybond polyvinyldiene difluoride membrane (GE healthcare) and submited to Edman degradation using a Procise proteins sequencer (Applied Biosystems, Weiterstadt, Germany). Scanning densitometry of SDS-PAGE were performed using the Quantity One software (Bio-Rad Laboratories). Concentration of protein bands corresponding to MycB were determined using standard solutions of thyroglobulin.

*2.9. ATP/Ppi exchange*

Substrate amino acid dependent ATP/PPi exchange reaction was measured as described elswhere [34] with some modifications. The reaction mixture consisted of 100 µl Tris-20 mM HCl pH 7.6, 1 mM MgCl$_2$, 5mM ATP, 1 mM tetra-sodium pyrophosphate, 1 µCi tetra-sodium [$^{32}$P] pyrophosphate, 0,5 mM asparagine and 20 µl of purified fractions. The samples were incubated for 40 min at room temperature. The [$^{32}$P]- labeled ATP was quantified by liquid scintillation methods.

## 3. Results

*3.1 Influence of the culture temperature on mycosubtilin production*

Mycosubtilin was measured after 72 h of culture, which correspond to the maximal level of production, at different temperatures for the *B. subtilis* wild-type strain ATCC6633 and BBG100, its constitutive mycosubtilin overproducing derivative [21]. As shown in Fig. 1a, the mycosubtilin production decreased in a temperature-dependent manner for both strains. Indeed, the mycosubtilin concentration in the culture broth decreased from 28 mg (mg DW)$^{-1}$ at 25 °C to 1 mg (mg DW)$^{-1}$ at 37 °C for the wild-type strain ATCC6633, whereas it decreased from 123 mg (mg DW)$^{-1}$ to 16 mg (mg DW)$^{-1}$ for BBG100 under the same conditions. For both strains, a regression analysis between the mycosubtilin concentration and the temperature led to a coefficient of determination ($R^2$) of 0,99 and 0,98 for ATCC6633 and BBG100, respectively. This demonstrated a direct relation between the mycosubtilin production and the growth temperature. Additional analysis on normalized values highlighted a temperature dependency of the mycosubtilin production equivalent whatever its level of production (Fig 1b).

*3.3. Influence of the culture temperature on myc expression*

The multienzymatic complex involved in mycosubtilin biosynthesis is encoded by the *myc* operon. To further characterize the influence of growth temperature on mycosubtilin production, a *myc-lacZ* reporter gene fusion was constructed and the expression of *myc* was determined at different temperatures. As little information is available on the *myc* promoter sequence, the intergenic region between the upstream *pbp* gene and *fenF*, the first gene of the *myc* operon was cloned and fused to a *spoVG-lacZ* promoter-less cassette.

Strain BBG117, a *B. subtilis* ATCC6633 derivative containing the *myc-lacZ* reporter fusion, was grown at different temperatures and the β-galactosidase activity was measured. Samples were collected at different time periods corresponding : (i) to the exponential growth phase, (ii) to the transition from the exponential to the stationary phase and (iii) to the stationary phase. As shown in Fig. 2, the level of β-galactosidase activities was maximal in the samples collected during the transition phase, suggesting that *myc* expression is maximal at that culture period. The expression of *myc* then decreased during the stationary phase. However, for all the temperatures tested, the β-galactosidase activity profiles were not significantly different, indicating that *myc* expression seems to be temperature-independent.

To further characterize the influence of the temperature on *myc* expression, the mRNA of *fenF*, the first ORF of the *myc* operon, was quantified in cold stress condition and compared to that obtained at 37°C. Therefore, mRNAs of *fenF* and *cspB*, a gene encoding the major cold shock protein, were quantified by RT-PCR using *rplL* mRNA, encoding the ribosomal protein L12, for data normalization. As shown in Fig. 3, inducting a cold shock response by growing culture of B. subtilis ATCC6633 at 20 °C resulted in a clear increase in the mRNA level of cspB. However, at this low temperature, no significant difference in the *fenF* mRNA level could be found as compared to the one observed at 37°C. This demonstrated that *myc* expression is not affected at low temperature.

*3.2 MycB purification and quantification*

The mycosubtilin synthetase is a modular multienzymatic complex with a relative molecular mass of more than 1300 kDa. In an attempt to estimate its relative abundance in *B. subtilis* cells, MycB, its larger subunit, was isolated and purified. This was performed for BBG100 grown at either 25 °C or 37 °C and for the negative control strain RFB136. The latter is a mycosubtilin non-producer derivative of *B. subtilis* ATCC6633 obtained by disruption of the *myc* operon. Cell-free extracts were prepared from cultures in late logarithmic phase of growth by lysozyme treatment, sonication and nucleic acid precipitation. Proteins were precipitated with ammonium sulfate and fractionated by size exclusion chromatography.

Fractions from BBG100 cultures were analysed by SDS-PAGE and compared to those obtained from RFB106. Protein bands with an apparent molecular mass of approximately 610 kDa were detected in fractions of BBG100 samples whereas they were not observed in RFB106 fractions (Fig. 4a). This apparent molecular mass is in good agreement with the 612 kDa deduced from the published *mycB* sequence, suggesting that the purified protein corresponds to MycB [5]. This was confirmed by the determination of the three first N-terminal amino acids of the 610 kDA protein (data not shown). The protein concentration of the purified enzyme was then estimated by scanning densitometry using a standard solution of thyroglobulin. As shown in Fig. 4 b, with values equal to 65 and 48 µg ml$^{-1}$, respectively, MycB concentration was slightly higher when BBG100 was cultured at 25 °C compared to that obtained at 37 °C.

To further characterise the influence of growth temperature on the intracellular mycosubtilin synthetase concentration, its abundance was estimated using one of the enzymatic reaction catalysed by MycB. In NRPS mechanisms, amino acids are first activated as an adenylate derivative in the A-domain of the synthetase before being incorporated into the peptidic chain. Among the three amino acids incorpored by MycB, L-asparagine is the sole specific to the mycosubtilin peptidic chain (*i.e.* not present in others lipopeptides). Therefore, its adenylation was measured by an ATP/PPi exchange assay in samples purified from BBG100 cultured at 25 and 37 °C and from RFB136 cultures used as a negative control. As shown in Fig. 4b, a slight increased level of asparagine adenylation could be observed in samples obtained from culture performed at 25 °C compared to that obtained at 37 °C. Indeed, value of ATP-PPi exchange were equal to 675 $10^3$ cpm and 486 $10^3$ cpm for samples obtained from cells cultured at 25 and 37 °C, respectively. For RFB136 samples, adenylation of L-asparagine was never detected (data not shown).

All of these observation demonstated that mycosubtilin synthetase could be somewhat more abundant in the cells cultured at 25 °C. However, those discrepancies are not sufficient to explain the 30-fold increased mycosubtilin production at low temperature.

*3.2. Mycosubtilin homologues production*

Changes in membrane lipid composition, and thus in the intracellular pool of fatty acids is one of the known consequences of the cell adaptation to low temperature [10]. This suggests that the proportion of the different mycosubtilin homologues, containing various chain length fatty acid moieties, could be affected by the growth temperature. Therefore, the mycosubtilin homologue proportions were measured after 72 h of growth for *B. subtilis* strains ATCC6633 and BBG100 cultured at 25, 30 and 37°C. As shown in Fig. 5, a significant increase in the proportion of mycosubtilin with odd-numbered fatty acids (i.e., C15 and C17) could be observed at lower temperature for both strains. By contrast, the proportion of the $C_{16}$ fatty acid homologue decreased under these conditions. Production of odd-numbered fatty acid homologues also seemed to be affected by the mycosubtilin production yield. Indeed, for the mycosubtilin overproducer BBG100, the C17 homologue represented 80% and 77% of the total mycosubtilin production at 25 and 30°C, respectively, whereas these proportions fall to 71 % and 63 % for the wild-type strain ATCC6633, under the same conditions (Table 2).

To confirm the increase in the production of mycosubtilin with long chain fatty acid at low temperature, we performed cold shock experiments. Both *B. subtilis* ATCC6633 and BBG100 strains were first cultured at 30°C for 8 h before being shifted to 25°C for 64 h. Lipopeptide production and homologue composition were then analysed and compared to those obtained for the cultures performed at 25 and 30°C for 72 h. As it was previously observed, significant differences in the proportion of the mycosubtilin homologues produced occurred for the cultures performed for 72 h at 25 and 30°C (Table 2). In addition, the differences in the C16 and C17 homologue proportions were more marked for the shifted culture than those obtained at 25°C (Table 2). Indeed, for BBG100, the C16/C17 homologue ratio were equal to 0.25 at 25°C and 0.12 for the shifted culture, whereas these ratio were 0.41 and 0.22 for the wild-type strain under the same conditions. This could result from a stronger response of the cell metabolism (*i.e*., a higher production of odd-numbered fatty acids) after cold stress. In all cases, the increase in odd-numbered homologues at 25 °C and after cold shock was mainly due to an increase in the synthesis of mycosubtilin with branched fatty acids moiety. Further cold shock experiments from 38 to 30°C performed with BBG100 yielded similar results (data not shown).

## 4. Discussion

In its natural environment, the surface layer of soil, *B. subtilis* is exposed to temperature fluctuations which induce modifications in its physiology and metabolism. At lower temperature, cells must face several problems, including low membrane fluidity, reduced enzyme activities, decreased initiation of translation due to stabilized secondary structures of mRNAs or slower protein folding [16]. Fatty acids are one of the most important building blocks of cellular materials. In *B. subtilis*, the membrane composition is characterized by a fatty acid profile dominated to a large extend by odd-numbered branched-chain fatty acids, with the major $C_{15}$ and $C_{17}$ species [16]. These latter were shown to play a major role in the correct physical state of the membrane lipids, which is required for optimal membrane structure and function. At lower temperature, the membrane fluidity must increase to avoid transition from a liquid crystalline into a gel-like phase state of the lipid bilayer. *Bacillus* cells respond to a decrease in the growth temperature by desaturating the fatty acids of their membrane lipids though the activation of the *Des* pathway [3, 8] and by increasing the proportion of ante-iso branched fatty acids which present a lower melting point [23, 33]. However, it was shown that the deletion of the *des* genes does not lead to any detectable phenotype after cold shock, indicating that *B. subtilis* rather adapts to low temperature by modifying its iso- and anteiso-fatty acid membrane composition [17]. These modifications must involve *de novo* fatty acid synthesis. Anteiso-methyl branches cannot be added by methylation of existing fatty acids, but are introduced as part of the primer molecule during the initiation of fatty acid synthesis. Anteiso-branched C15 and C17 fatty acids are formed from α-keto-β-methylvalerate and 2-methyl-butyryl-CoA, which both derive from isoleucine. The modification of the membrane composition at low temperature were demonstrated in *B.*

*subtilis* JH642 [17]. These results suggested a *de novo* synthesis of anteiso C15:0 and C17:0 fatty acids from isoleucine or threonine present in the culture medium as a response to a cold stress.

In the first step of the mycosubtilin synthesis pathway, the acyl CoA-ligase domain of *myc* couples coenzyme A to a cytoplasmic long chain fatty acid. The activated fatty acid is then transferred to the 4-phosphopantetheine cofactor of the first acyl carrier domain of the mycosubtilin synthetase. In the subsequent reactions of condensation, adenylation and thiolation reactions, the mycosubtilin molecule is synthetized analogously to other non-ribosomal peptides [5]. It was also recently demonstrated by pulse-chase experiments that some flexibility exists in the length of fatty acid group incorporated in mycosubtilin molecules [13]. Taken together, all of these findings suggest that the predominant fatty acid homologue present in the cytoplasmic pool compatible with the lipopeptide structure is preferentially incorporated. Therefore, the variation in the composition of this pool following a temperature modification should also influence the production of mycosubtilin in the same manner. Our results are clearly supportive with this hypothesis since both low growth temperature and cold shock, favouring *de novo* synthesis of odd-numbered fatty acids, led to an increased production of mycosubtilin with a $C_{15}$ and $C_{17}$ fatty acid moiety.

Low growth temperature was also shown to significantly increase the mycosubtilin production yield. Our results suggest that this increase was not due to an overexpression of *myc* at low temperature since (i) *myc* induction and the concentration of the resulting mRNA were not modified at low temperature; (ii) the increase of mycosubtilin production was also observed for the constitutive producer BBG100. Therefore, this increased level of production could rather result from a higher activity of the mycosubtilin synthetase at low temperature. This hypothesis is reinforced by the fact that the differences in the mycosubtilin synthetase concentration in the cells cultured at different temperature are not sufficient to explain the difference in mycosubtilin production yield. This overproduction at low temperature is consistent with previous findings observed for *B. subtili*s RB14 in solid-state fermentation for another member of the iturin family, iturin A [24, 26].

Lipopeptides were reported to be a key parameter in biofilm formation and rhizosphere colonization [27, 28]. Therefore, the production of the most active homologue in large amount facilitates these phenomena. When considering that the temperature in the rhizosphere is rather close to 20 than 37 °C [9], it is not surprising that the lipopeptide production mechanism seems to be adapted to this low temperature.

**Acknowledgements**

This work received financial supports from the Université des Sciences et Technologies de Lille, the Région Nord-Pas-de-Calais and the Fonds Européen pour le Développement de la Recherche. P. Fickers was a recipient of a Post-Doc grant from the Université des Sciences et Technologies de Lille and is now a Postdoctoral Researcher at the F.N.R.S. (Fonds National de la Recherche Scientifique). We are grateful to B. Wathelet (Unité de Chimie Biologique Industrielle, Faculté des Sciences Agronomiques de Gembloux, Gembloux, Belgium) for performing MALDI-TOF measurement.

**Figures legends.**

**Fig. 1** : Mycosubtilin productivity by *B. subtilis* ATCC6633 (♦) and BBG100 (■) after 72 h of growth in Landy medium at 25, 30 and 37 °C. Raw values (A) and normalized values (B). Results are mean values of three independent experiments. Standard deviations were less than 10 % of average value.

**Fig. 2**. β-galactosidase activity determined for *B. subtilis* BBG117 after 4, 6 and 8 h of growth in Landy medium at 25, 30 and 37 °C. Results are mean values of three independent experiments. Standard deviations were less than 10 % of average value.

**Fig. 3**. Expression of *rplL*, *fenF* and *cspB* at 20 and 37 °C. Electrophoretic profile of purified mRNAs subjected to RT-PCR (A). MTM values of *fenF* and *cspB* gene expression obtained after scanning densitometry and correction for the level of *rplL* for a given sample (B). Displayed data are one representative result of two independent experiments.

**Fig. 4**. Purification and caracterisation of MycB in cell-free extract of BBG100 and RFB136 strains cultured in Landy medium at 25 °C and 37 °C. (A) Coomassie-stained 5 % SDS polyacrylamide gel showing the protein composition of the fraction obtained after size exclusion chromatography and tested for adenylation reaction. Five μg of protein were loaded per lane. Fraction showing the most intense 610 kDa band are presented for BBG100 whereas the corresponding chromatographic fraction is presented for RFB136 (B) Determination of the MycB protein concentration in the corresponding fraction by scanning densitometry (grey) and ATP/PPi exchange assay (black). Displayed data are one representative result of two independent experiments.

**Fig. 5**. Relative abundance of the $C_{15}$ (black), $C_{16}$ (white) and $C_{17}$ (grey) mycosubtilin homologues determined for *B. subtilis* ATCC6633 (A) and the mycosubtilin overproducing strain BBG100 (B) after 72 h of growth in Landy medium. The data are mean values of three independent experiments. All the values of mycosubtilin homologue production were significantly different for a given strain according to a Student's *t*-test at P<0.05.

Table 1

Plasmids and strains used in this study

| Plasmids | Description, structure or sequence (5'-3') | Source, restriction site |
|---|---|---|
| pGEM-T Easy | PCR fragments cloning vector | Promega |
| pDG1661 | Integrative vector at *amyE* locus | [11] |
| pBG107 | P*myc* promoter into pGEM-T Easy | This work |
| pBG111 | P*myc* –*lacZ* into pDG1661 | This work |
| RFP104 | *Isce*I-Kan^R –*Isce*I fragment into pGEM-T Easy | Fickers et al unpublished results |
| RFP119 | *mycA* PT cassette into pGEM-T Easy | This work |
| RFP120 | Kan^R fragment into RFP119 at *ISce*I site | This work |

| Primers | | |
|---|---|---|
| pMYCfo | CGTCAA<u>GAATTC</u>TTTATCATTCCATATATACG | *Eco*RI |
| pMYCrev | ATTCATT<u>GGATCC</u>CTCCAATCTTTTCGAACGG | *Bam*HI |
| AmyEfo | GGAAGCGGAAGAATGAAGTAAGAGGG | |
| AmyERev | GCCAGGCTGATTCTGACCGGGCAC | |
| FenFfo | CAAAATGCAGATCCTGAGCA | |
| FenFrev | GGCATAGTCATGTGCGTTTG | |
| rplL fo | GCTTCCGTTAAAGAAGCAACTG | |
| RplLrev | AGAAGCGCCAACTTCTTCAA | |
| cspB Fo | AAAAGGTTTCGGATTCATCG | |
| cspBrev | AACGTTAGCAGCTTGTGGTC | |
| RFO120 | GGAGTCAGCTCGGGGTATATGTAGGG | |
| RFO121 | <u>ATTACCCTGTTATCCCTA</u>ACCCAACAGCGATTCCGTTCAATTGG | I-*Sce*I |
| RFO122 | <u>TAGGGATAACAGGGTAAT</u>GGACGCAACTGCTTCTTGTCTACGGCC | I-*Sce*I |
| RFO123 | ATAATACGCATATCCGGCATTTGATCGTGT | |

*B. subtilis* strains

| | | |
|---|---|---|
| ATCC 6633 | Wt | Lab stock |
| BBG100 | ATCC6633 *myc* controlled by *repU* promoter | [21] |
| BBG117 | ATCC 6633 *amyE*::P*myc-lacZ* | This work |
| RFB136 | ATCC 6633 *mycA*::Kan$^R$ | This work |

Table 2

Repartition of the C16 and C17 mycosubtilin homologues produced by *B. subtilis* ATCC6633 cultured for 72 h at 25, 30°C and during temperature shift experiments. B and L are for branched and linear isomers respectively whereas Tot are for the total amount detected. ND : non-determined. All the values for a given strain were significantly different for each conditions tested according to a Student's *t*-test at P<0.05

| Temperature | ATCC6633 | | BBG100 | |
|---|---|---|---|---|
| | C16 (%) | C17 (%) | C16 (%) | C17 (%) |
| | Tot (B/L) | Tot (B/L) | Tot (B/L) | Tot (B/L) |
| 25 | 29 (ND) | 71 (63/8) | 20 (ND) | 80 (66/14) |
| 30 | 37 (ND) | 63 (53/10) | 23 (ND) | 77 (57/20) |
| Down-shift | 18 (ND) | 80 (69/11) | 11 (ND) | 89 (73/16) |

Figure 1

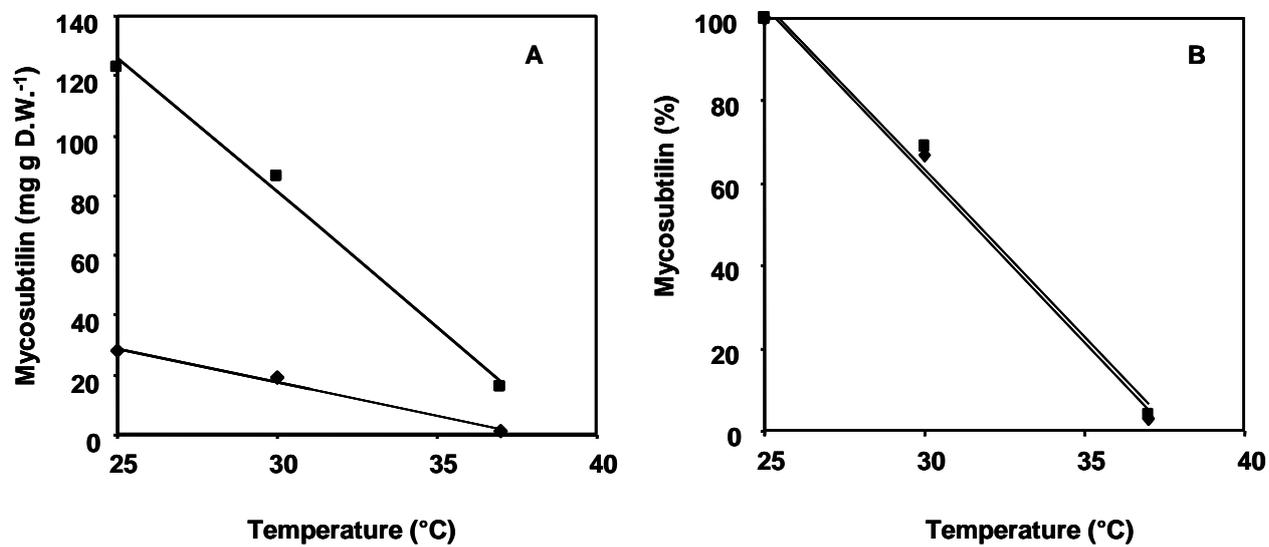

Figure 2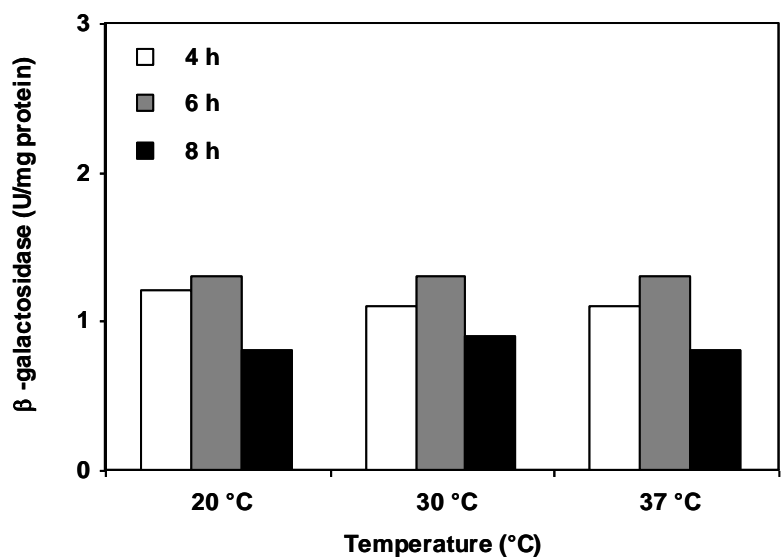

**Figure 3**

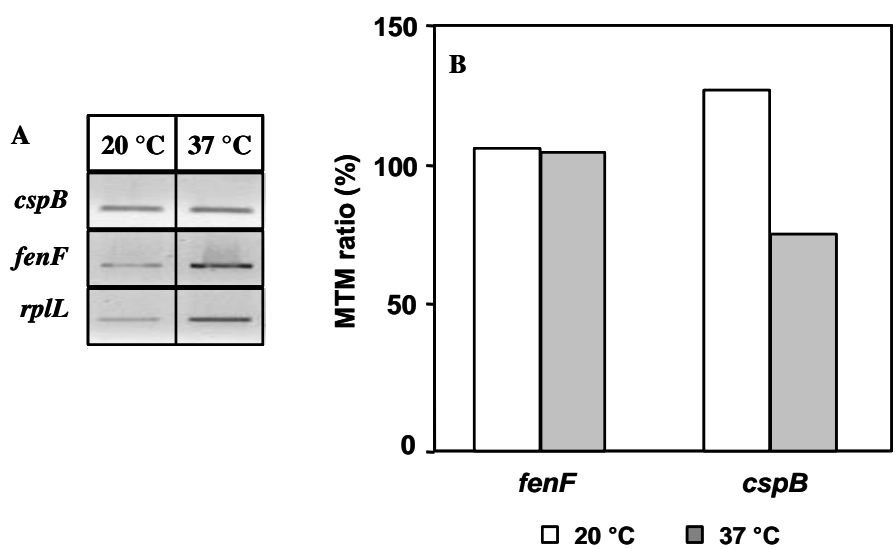

**Figure 4**

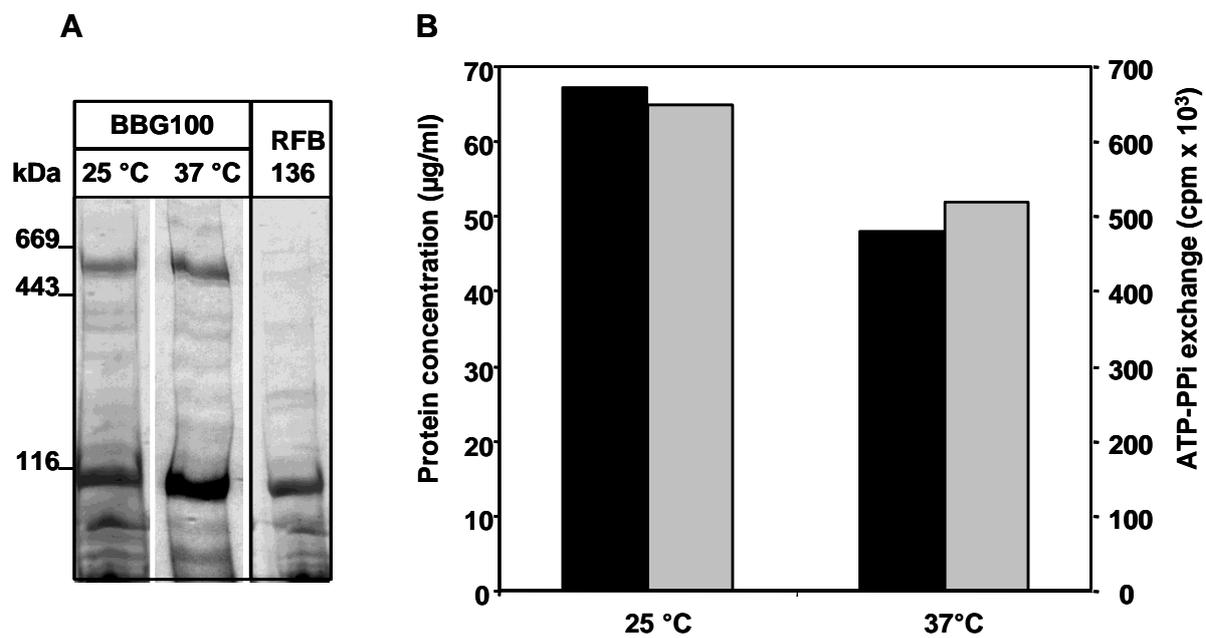

Figure 5

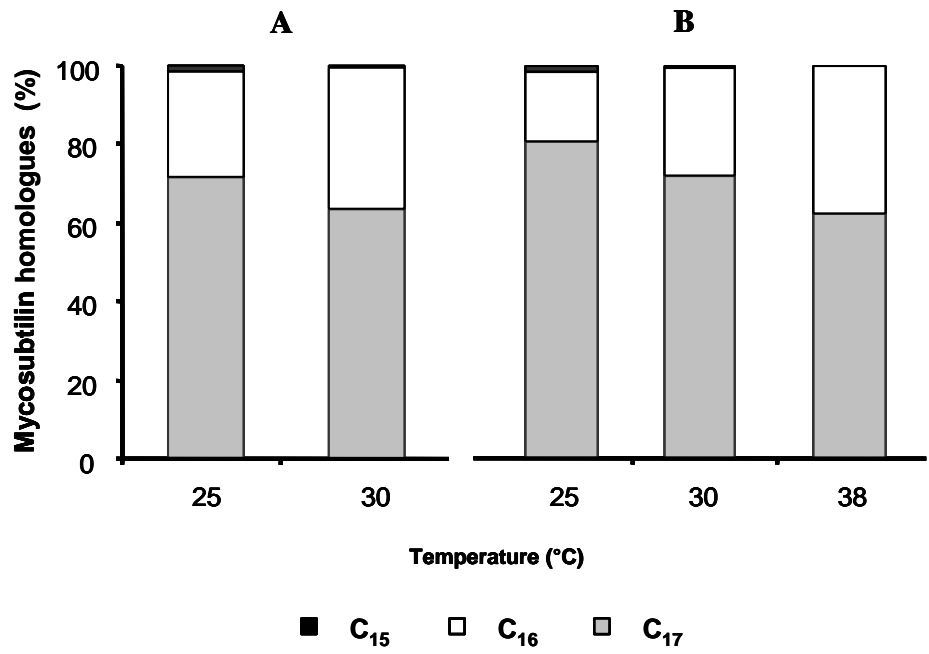